\documentclass[twoside]{article}
\usepackage{fleqn,psfig,here,espcrc2}

\title{\vspace{-4.0cm}
       \rightline{\normalsize KEK-CP-094}
       \vspace{-0.1cm}
       \rightline{\normalsize HUPD-9918}
       \vspace{2.9cm}
  $O(\alpha_{s}a)$ matching coefficients for axial vector current
  and $\Delta B$$=$2 operator\thanks{Poster presented by K-I. Ishikawa.}
}

\author{
  K-I. Ishikawa
  \address{ High Energy Accelerator Research Organization (KEK),
             Tsukuba, Ibaraki 305-0801, Japan}
  \thanks{K-I.I. is supported by the JSPS Research Fellowship.},
  T. Onogi
  \address{ Department of Physics, Hiroshima
             University, Higashi-Hiroshima, Hiroshima 739-8526, Japan}
  \thanks{T.O. is supported by the Grants-in-Aid of the Ministry
          of Education (No. 10740125).},
  N. Yamada$^{\mbox{\scriptsize\ b}}$
}

\begin{document}

\begin{abstract}
  We present a calculation of the perturbative matching coefficients 
  including mixing with higher dimensional operators for the temporal 
  component of the heavy-light axial current, $A_{4}$, and the $\Delta B=2$ 
  operator, $O_S$. For $A_{4}^{\mbox{\scriptsize static, NRQCD}}$,
  calculations with various RG-improved gauge actions are peformed.
  Matching coefficients with NRQCD 
  and heavy-clover actions are also compared. 
\vspace*{-1em}
\end{abstract}

\maketitle

\section{Introduction}
It is known that there is a large mixing of lattice heavy-light 
axial current with higher dimensional operators in the static limit 
and in NRQCD~\cite{MIXING,FB}, which gives significant decrease of $f_B$. 
However, not much is known about the mixing effect for more general cases, 
even though lattice computations of various matrix elements 
in heavy-light system have been studied from phenomenological interest.

In this work, we compute the mixing coefficients with higher 
dimensional operators in lattice theory for various cases
of practial use in heavy-light system, i.e:
(1) for $A_4$ with RG-improved  gauge action
(2) for $A_4$ with heavy-clover action in Fermilab interpretation~\cite{EKM},
and (3) for $O_S$ which is needed to extract the width difference 
of $B_{s}$ meson.

\section{Matching of $A_{4}$ with improved gauge actions}
Improved action has advantages in simulations with dynamical quarks,
when one is forced to use coarse lattices due to the lack of 
computational power. For example,
CP-PACS collaboration has started calculation of $f_B$ with 
dynamical quarks using Iwasaki-action~\cite{CPPACS}.
However, until now the 1-loop computation of heavy-light axial vector
current did not exist. We compute the matching coefficient of heavy-light 
axial current $A_{4}$ for practical application to
CP-PACS~\cite{CPPACS}  and
for studying the effect of improvement on the matching coefficients.

We employ the static and the $O(1/M)$ NRQCD for heavy quarks,
and the massless clover action for light quarks.
The gauge action is described as
\begin{eqnarray}
 S_{\mbox{\rm\scriptsize gluon}}&\!\!\!\!=&\!\!\!\!
\frac{1}{g^{2}}\biggl[
 c_{0}\!\!\!\!\!\!
 \sum_{\mbox{\rm\scriptsize plaquette}}\!\!\!\!\!
 \mbox{\rm Tr}U_{\mbox{\rm\scriptsize pl}}
+c_{1}\!\!\!\!\!\!
 \sum_{\mbox{\rm\scriptsize rectangle}}\!\!\!\!\!
 \mbox{\rm Tr}U_{\mbox{\rm\scriptsize rtg}}  
 \nonumber \\
&&
+c_{2}\!\!
 \sum_{\mbox{\rm\scriptsize chair}}\!
 \mbox{\rm Tr}U_{\mbox{\rm\scriptsize chr}}
+c_{3}\!\!\!\!\!\!\!\!\!\!\!
 \sum_{\mbox{\rm\scriptsize parallelogram}}\!\!\!\!\!\!\!\!\!\!
 \mbox{\rm Tr}U_{\mbox{\rm\scriptsize plg}}
\biggr].
\end{eqnarray}
At the 1-loop level, the choice of the gluon action is
specified by $c_1$ and $c_{23}$$\equiv$$c_{2}$$+$$c_{3}$.
We adopt the following five combinations for the gauge parameters;
plaquette: $(c_{1}, c_{23})$$=$$(0, 0)$,
tree-level improved: $(-1/12, 0)$,
RG-improved:  $(-0.331, 0)$,
              $(-0.27, -0.04)$, and
              $(-0.252, -0.17)$.
The combination $(-0.331, 0)$ corresponds to Iwasaki action.

\begin{table*}[t]
\begin{center}
{\small
\setlength{\tabcolsep}{1mm}
\begin{tabular}{cccccccccc}\hline
\multicolumn{2}{c}{gluon action} & & & & & & & & \\
\cline{1-2} 
$c_{1}$&$c_{23}$
&\raisebox{1.5ex}[0pt]{$R_{0}^{(0)}$}
&\raisebox{1.5ex}[0pt]{$R_{1}^{(0)}$}
&\raisebox{1.5ex}[0pt]{$R_{0}^{\mbox{\rm\tiny (disc)}}$}
&\raisebox{1.5ex}[0pt]{$R_{1}^{\mbox{\rm\tiny (disc)}}$}
&\raisebox{1.5ex}[0pt]{$Z_{h}$}
&\raisebox{1.5ex}[0pt]{$Z_{l}$}
&\raisebox{1.5ex}[0pt]{
\begin{tabular}{c}
\hspace{-1em} $\rho_{\gamma_5\gamma_4}^{(0)}$ \hspace{-1em}\\
\hspace{-2em} $-(1/\pi)\ln(am_b)$ \hspace{-1em} 
\end{tabular}
}
&\raisebox{1.5ex}[0pt]
{$\rho_{\gamma_5\gamma_4}^{\mbox{\rm\tiny (disc)}}$}  \\ \hline
    0 & 0    & 0.57976(6)&-0.32709(1)&-0.54719(6)&0.48795(11)
             & 0.4807(3) & 0.1814(2) 
             &-1.3175(2) & 1.0351(1)\\
-1/12 & 0    & 0.56673(6)&-0.26963(1)&-0.44345(6)&0.33315(10)
             & 0.4175(3) & 0.0550(3)
             &-1.1522(2) & 0.7766(1)\\
-0.331& 0    & 0.54788(5)&-0.18085(1)&-0.30145(5)&0.03856(7)
             & 0.2022(3) &-0.1311(3)
             &-0.8439(2) & 0.3400(1)\\
-0.27 &-0.04 & 0.54986(5)&-0.18711(1)&-0.31378(5)&0.06723(7)
             & 0.2021(3) &-0.1185(3)
             &-0.8583(2) & 0.3810(1)\\
-0.252&-0.17 & 0.54750(5)&-0.16757(1)&-0.29003(5)&-0.00133(7)
             & 0.0910(2) &-0.1602(3)
             &-0.7600(2)& 0.2887(1)\\ \hline
\end{tabular}
}
\caption{Perturbative coefficients for static(heavy)-clover(light) currents
with improved gluon action. $Z_{l}$ are tadpole improved with $\kappa_{c}$.}
\label{STATICLIGHT}
\end{center}
\vspace*{-2em}
\end{table*}

In the static case, the continuum heavy-light bilinear 
operators ($\overline{\mbox{MS}}$ scheme)
$J^{\mbox{\rm\scriptsize cont}}_{\Gamma}$ with arbitrary Dirac matrix 
$\Gamma$  are expressed 
by the lattice operators to $O(\alpha_s a)$ as
\begin{eqnarray}
J^{\mbox{\rm\scriptsize cont}}_{\Gamma}&\!\!\!\!\!\! = &\!\!\!\!\!\!
  \left[1\!+\!\alpha_{s}\rho_{\Gamma}^{(0)}\right]
\!  J^{\mbox{\rm\scriptsize latt} (0)}_{\Gamma}
\!\!+ \!
\alpha_{s}\rho_{\Gamma}^{\mbox{\rm\scriptsize (disc)}} 
\!  J^{\mbox{\rm\scriptsize latt (disc)}}_{\Gamma}\!\!\!\!\!,
\label{eqn:STATIC}
\end{eqnarray}
where $J^{\mbox{\rm\scriptsize latt (0)}}_{\Gamma}$$\equiv$$\bar{q}\Gamma b$
is the leading lattice operator
and 
$J^{\mbox{\rm\scriptsize latt (disc)}}_{\Gamma}$$\equiv$$(\bar{q}\vec{\gamma}\cdot\!\!\stackrel{\leftarrow}{aD})\Gamma b$
is the higher dimensional operator of $O(a)$. 
$q$ is the massless clover quark field, 
and $b$$\equiv$$(Q, 0)^{T}$ with the static two component quark field $Q$.
The 1-loop coefficients in Eq.~(\ref{eqn:STATIC}) are
expressed as 
\begin{eqnarray}
\hspace{-2em}
&&
  \rho_{\Gamma}^{\mbox{\rm\scriptsize (0)}}\!\!=\!
  C^{\mbox{\scriptsize cont}}_{\Gamma}
   -\left[ R_{0}^{(0)} +R_{1}^{(0)}G 
              +\frac{1}{2} \left(Z_{l}+Z_{h}\right)
    \right]\!, \nonumber\\
\hspace{-2em}
&&
   \rho_{\Gamma}^{\mbox{\rm\scriptsize (disc)}} \!\!=
  \! \frac{r( c_{\mbox{\scriptsize sw}}\! -\!1)}{3\pi}
     \ln\! \left(a^{2}\lambda^{2}\right)
\!-\!\left(    R_{0}^{\mbox{\rm\scriptsize (disc)}} 
     \!\!+\!\! R_{1}^{\mbox{\rm\scriptsize (disc)}}G\right)\!,\nonumber
\end{eqnarray}
with known continuum 1-loop correction
$C^{\mbox{\scriptsize cont}}_{\Gamma}$ (see~\cite{STATICFOURFERMIMIXING}).
Here $G$$=$$\gamma_0\Gamma\gamma^{0}$, and
$\lambda$ is a gluon mass introduced to regularize the infrared divergence,
which is cancelled by setting $c_{\mbox{\scriptsize sw}}$$=$$1$.
$R^{\mbox{\scriptsize (0, disc)}}_{0,1}$'s are the vertex correction,
$Z_{l}$ and $Z_{h}$ are the wave function renormalization
of light and static quarks in the lattice theory, respectively.

The numerical values for $R^{\mbox{\rm\scriptsize (0, disc)}}_{0,1}$,
$Z_{l,h}$, and $\rho_{\gamma_5\gamma_4}^{\mbox{\rm\scriptsize (0, disc)}}$
are shown in Table~\ref{STATICLIGHT}.
We find that
the coefficients  for the tree-level improved (RG-improved) gauge action(s)
are reduced by 20-30\% (40-70\%) from those for the plaquette action.
This is encouraging for simulations on a coarse lattice,
since it suggests that the perturbative error in the matching coefficients 
with improved actions are much smaller than those with the unimproved action,
assuming $\alpha_V$'s with the same lattice spacing $a$ have almost 
the same values. 

We have calculated the matching coefficients for $A_4$ with
the NRQCD and Iwasaki actions.
The relation between the continuum and NRQCD 
operators to $O(\alpha_s a)$ and $O(\alpha_s/M)$ is given by
\begin{equation}
A_{4} \!\! =\!\!
\left[1+\alpha_{s}\rho_{A}^{(0)}\right] J^{(0)}_{4} 
\!\! +\! \alpha_{s}\rho_{A}^{(1)} J^{(1)}_{4}
\!\! +\! \alpha_{s}\rho_{A}^{(2)} J^{(2)}_{4}\!\!,
\label{eqn:MSNRQCD}
\end{equation}
where $J^{(0)}_{4}$$\equiv$$\bar{q}\gamma_{5}\gamma_{4}b$,
      $J^{(1)}_{4}$$\equiv$$- \bar{q}\gamma_{5}\gamma_{4}
                             (\vec{\gamma}\!\cdot\!\vec{D}b)/2M_{0}$,
      $J^{(2)}_{4}$$\equiv$$ (\bar{q}\vec{\gamma}
                              \cdot\!\!\!\!\!\stackrel{\leftarrow}{D})
                              \gamma_{5}\gamma_{4}b/2M_{0}$, and
$b$$\equiv$$[1$$-$$\vec{\gamma}\!\!\cdot\!\!\vec{D}/2M_{0}]$
$(Q, 0)^{T}$,
with the NRQCD two component field $Q$ and the bare heavy qurak
mass $M_{0}$.
We used the same NRQCD action
as that in Ref.~\cite{MIXING}, where details of the computation 
with plaquette action is presented.

Figure~\ref{fig:NRPLVSRG} shows the quark mass dependence
of the 1-loop coefficients with Iwasaki (plaquette) gauge action
by filled (open) symbols.
$\rho^{(2)}_{A}/2aM_{0}$ has nonzero value in the static limit, and 
$\alpha_s\rho^{(2)}_{A}J^{(2)}_{4}$ contains $O(\alpha_s a)$ correction.
As was observed in the static case,
the matching coefficients for Iwasaki action (filled symbols) are
much smaller than those for plaquette one (open symbols). 
It is therefore expected that the decrease of $f_B$ by the mixing
with higher dimensional operators is not significant
for RG-improved gauge action compared to the case with 
plaquette action~\cite{CPPACS}.

\section{Matching of $A_{4}$ with heavy-clover action}
We now turn to the studies on the effect of the heavy quark 
action on the matching coefficients of $A_{4}$.
We compare the result with clover action to that with NRQCD action
where the gauge action is plaquette one.

The relation between continuum $A_{4}$ and lattice counterparts
with heavy-clover quarks
is the same as Eq.~(\ref{eqn:MSNRQCD}) except for
the lattice heavy quark field $b$ and
the bare quark mass. The heavy quark field $b$ and  
the quark mass $aM_{0}$ are replaced by
$b$$=$$[1 + d \vec{\gamma}\!\cdot\!\vec{aD}]\psi_{\mbox{\rm\tiny CL}}$
with the four component spinor field $\psi_{\mbox{\rm\tiny CL}}$
of clover quarks and the bare kinetic mass $aM_2$, respectively.
In order to reproduce the same infrared divergence in the lattice
theory as in the continuum one, $d$$\equiv$$aM_{0}/2(1+aM_{0})(2+aM_{0})$
and $c_{\mbox{\scriptsize sw}}$$=$$1$ for both of the heavy and light clover quarks 
are required.

Figure~\ref{fig:CLVSNR} shows the matching coefficients 
with clover action (filled symbols)
together with the result with NRQCD action (open symbols) against
$1/aM_{2 (0)}$.
For heavier quark mass, the coefficients for clover
action have similar behavior to the NRQCD case and approach
the same static limit.
In the region of $1/aM_{2}$$\sim$0.4-1, the
coefficients $\rho^{(1)}_{A}$ and $\rho^{(2)}_{A}$ have almost the same
magnitude with oposite sign, which means that the contribution from 
$J^{(1)}_{4}$ partly cancels that from $J^{(2)}_{4}$ in calculating 
$f_{B}$.
Therefore, it suggests that the 1-loop mixing is small 
for $f_{B}$ with  heavy-clover quarks in contrast to NRQCD 
on a lattice with 4.2$>$$a^{-1}$$>$1.7 GeV assuming $M_{2}$$=$4.2 GeV.
This is consistent with the studies
of $f_{B}$ with heavy-clover action~\cite{FBCLOVER},
in which no significant lattice spacing dependence was observed.

\newcommand{\figwidth}{2.00in}
\begin{figure}[t]
\begin{center}
\leavevmode\hspace*{-1.0em}
\psfig{file=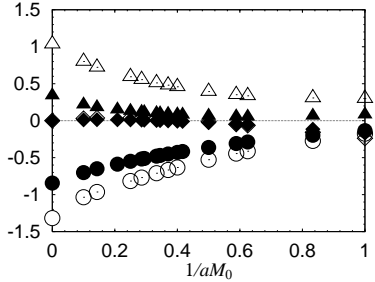,width=\figwidth}
\vspace*{-3.5em}
\caption{1-loop coefficients
for $A_{4}$ with NRQCD and 
Iwasaki (plaquette) actions as filled (open) symbols.
{\Large$\circ$}'s: $\rho_{A}^{(0)}$ without $log(aM_{0})$;
{\Large$\diamond$}'s: $\rho_{A}^{(1)}/2aM_{0}$;
{\small$\triangle$}'s: $\rho_{A}^{(2)}/2aM_{0}$.
Results at $1/aM_{0}$$=$$0$ are calculated with the static action.}
\label{fig:NRPLVSRG}
\end{center}
\vspace*{-3.0em}
\end{figure}
\begin{figure}[t]
\vspace*{-1em}
\begin{center}
\leavevmode\hspace*{-1.0em}
\psfig{file=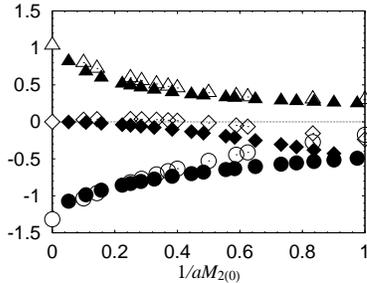,width=\figwidth}
\vspace*{-3.5em}
\caption{Same as Figure~\ref{fig:NRPLVSRG},
except for clover (NRQCD) with plaquette actions.
Filled (open) symbols : clover (NRQCD).}
\label{fig:CLVSNR}
\end{center}
\vspace*{-3em}
\end{figure}

\section{Matching of $O_S$}
We have calculated the matching coefficients for $O_S$ with
the lattice static and massless clover quarks including $O(\alpha_{s}a)$
correction employing the plaquette gauge action.
The operator identity between continuum $O_{S}$ and lattice
counterparts is obtained as
\label{sec:OLOS}
\begin{eqnarray}
\hspace{-2em}
&&    O_S^{\rm con}\!\!=\!\!
  \left[ 1 + \frac{\alpha_s}{4 \pi}
          \left\{   \frac{4}{3}\ln\left(a^2 m_b^2\right)
                  + \frac{16}{3}\ln\left(\frac{\mu^2}{m_b^2}\right)
          \right.
  \right. \nonumber \\
\hspace{-2em}
&&\hspace{1em}
    - 3.86 \biggr\}\biggr]\! O_S^{\rm lat}
  \!+\! \frac{\alpha_s}{4 \pi} \left[  0.77 \right]\! O_P^{\rm lat}
  \!+\! \frac{\alpha_s}{4 \pi} \left[  0.13 \right]\! O_R^{\rm lat}
\nonumber \\
\hspace{-2em}
&&
  + \frac{\alpha_s}{4 \pi}\!
    \left[\! -\! \frac{2}{3}\ln\left(a^2 m_b^2\right)
          \! +\! \frac{1}{3}\ln\left(\frac{\mu^2}{m_b^2}\right)
          \! +\! 3.91
    \right]\! O_L^{\rm lat} 
\label{eqn:STATICOS}
\\
\hspace{-2em}
&&
    +   \frac{\alpha_s}{4 \pi}\! \left[ -6.88 \right]\! O_{SD}^{\rm lat}
  \!+\! \frac{\alpha_s}{4 \pi}\! \left[  2.58 \right]\! O_{LD}^{\rm lat} 
  \!+\! \frac{\alpha_s}{4 \pi}\! \left[  1.15 \right]\! O_{PD}^{\rm lat},
 \nonumber
\end{eqnarray}
where
\begin{eqnarray}
\hspace{-2em}
&&
O_P =   2 \left[ \overline{b}\gamma_\mu P_L q\right]
            \left[ \overline{b}\gamma_\mu P_R q\right]
        +12 \left[ \overline{b} P_L q\right]
            \left[ \overline{b} P_R q\right],\nonumber\\
\hspace{-2em}
&&
O_{SD}\!\!= \left[ \overline{b} P_L q\right]
            \left[ \overline{b} P_L
            (a \vec{D}\!\cdot\!\vec{\gamma}q)\right],\nonumber\\
\hspace{-2em}
&&
O_{PD}\!\!=\!2\!\left[ \overline{b}\gamma_\mu P_L q\right]\!\!
                \left[ \overline{b}\gamma_\mu P_R
                (a \vec{D}\!\cdot\!\vec{\gamma}q)\right] \nonumber \\
\hspace{-2em}
&&\hspace{4em}
          +12\! \left[ \overline{b} P_L q\right]\!\!
                \left[ \overline{b} P_R
                (a \vec{D}\!\cdot\!\vec{\gamma}q)\right].
\end{eqnarray}
The definition of $O_{S,L,R,LD}$'s and
details of the calculation are the same as those in 
Refs.~\cite{STATICFOURFERMIMIXING}.

The $O(\alpha_{s}a)$ contributions appear as
$O_{SD,LD,PD}$ in Eq.~(\ref{eqn:STATICOS}).
Using the vacuum saturation approximation, we roughly estimate
the $O(\alpha_{s}a)$ correction at $\sim$20-40\%
for $\langle O_{S}\rangle$ and $\sim$5-10\% for $B_{S}$ with
$\alpha_V$ at $\beta$$\sim$5.7-6.1.
A detailed study on the operator mixing with the actual simulation
data is desired.



\begin{thebibliography}{99}
\bibitem{MIXING}
C.J.~Morningstar and J.~Shigemitsu, Phys. Rev. D {\bf 57} (1998) 6741.

\bibitem{FB}
A. Ali Khan {\it et al.},Phys. Lett. {\bf B427} (1998) 132;
J. Hein, Nucl. Phys. {\bf B}(Proc. Suppl.) {\bf 63A-C} (1998) 347;
JLQCD Collaboration: K-I. Ishikawa {\it et al.}, hep-lat/9905036.

\bibitem{EKM}
A. X. El-Khadra, A.S. Kronfeld and P.B. Mackenzie,
Phys. Rev. D {\bf 55} (1997) 3933.

\bibitem{CPPACS}
CP-PACS collaboration, presented by A.~Ali~Khan, these proceedings.

\bibitem{STATICFOURFERMIMIXING}
K-I. Ishikawa, T. Onogi, and N. Yamada, Phys. Rev. D {\bf 60} (1999)
034501 and references therein.

\bibitem{FBCLOVER}
JLQCD collaboratoin: S. Aoki {\it et al.}, Phys. Rev. Lett. {\bf 80}
(1998), 5711; A.X. El-Khadra {\it et al.}, Phys. Rev. {\bf D 58}
(1998) 014506.

\end{thebibliography}
\end{document}